\documentclass[11pt]{article}
\usepackage{amssymb,latexsym,amsmath,amsbsy}
\usepackage[dvips]{graphicx}
\usepackage{cite}
\newtheorem{theo}{Theorem}

\newtheorem{coro}[theo]{Corollary}

\newtheorem{prop}[theo]{Proposition}
\def\nn{\nonumber}

\def\ch{\mathop{\rm char}\nolimits}

\def\qdots{\mathinner{\mkern1mu\raise1pt\vbox{\kern7pt\hbox{.}}\mkern2mu
 \raise4pt\hbox{.}\mkern2mu\raise7pt\hbox{.}\mkern1mu}}
\def\Z{{\mathbb Z}}

\def\C{{\mathbb C}}
\def\gl{\mathfrak{gl}}

\def\u{\mathfrak{u}}
\def\h{\mathfrak{h}}
\def\so{\mathfrak{so}}
\def\sp{\mathfrak{sp}}
\def\osp{\mathfrak{osp}}

\begin{document}
\begin{center}
{\Large \bf
The parafermion Fock space and explicit $\so(2n+1)$ representations} \\[5mm]
{\bf N.I.~Stoilova}\footnote{E-mail: Neli.Stoilova@UGent.be; Permanent address:
Institute for Nuclear Research and Nuclear Energy, Boul.\ Tsarigradsko Chaussee 72,
1784 Sofia, Bulgaria} {\bf and J.\ Van der Jeugt}\footnote{E-mail:
Joris.VanderJeugt@UGent.be}\\[1mm]
Department of Applied Mathematics and Computer Science,
Ghent University,\\
Krijgslaan 281-S9, B-9000 Gent, Belgium.
\end{center}


\begin{abstract}
The defining relations (triple relations) of $n$ pairs of parafermion operators $f_j^\pm$ 
($j=1,\ldots,n$) are known to coincide with a set of defining relations for the Lie algebra
$\so(2n+1)$ in terms of $2n$ generators. With the common Hermiticity conditions, this means
that the ``parafermions of order~$p$'' correspond to a finite-dimensional unitary irreducible representation
$W(p)$ of $\so(2n+1)$, with highest weight $(\frac{p}{2}, \frac{p}{2},\ldots,\frac{p}{2})$.
Although the dimension and character of $W(p)$ is known by classical formulas, there is no
explicit basis of $W(p)$ available in which the parafermion operators have a natural action.
In this paper we construct an orthogonal basis for $W(p)$, and we present the explicit actions
of the parafermion generators on these basis vectors. We use group theoretical techniques,
in which the $\u(n)$ subalgebra of $\so(2n+1)$ plays a crucial role: a set of Gelfand-Zetlin
patterns of $\u(n)$ will be used to label the basis vectors of $W(p)$, and also in the
explicit action (matrix elements) certain $\u(n)$ Clebsch-Gordan coefficients are essential.
\end{abstract}

\vskip 20mm

\setcounter{equation}{0}
\section{Introduction} \label{sec:Introduction}%

It has been known for a long time that Fermi-Dirac and Bose-Einstein statistics do
not yield all possible descriptions of particle systems~\cite{Green, Greenberg}.
The classical work of Green~\cite{Green} is nowadays considered as the basic 
underlying mathematical formulation of the problem of generalized quantum statistics.
In this paper, the notions of Bose oscillators or bosons and Fermi oscillators or fermions
are generalized to parabosons and parafermions.
Parabosons and parafermions are of interest in many applications, in particular in quantum field 
theory~\cite{Haag, Ohnuki, Wu} and generalizations of quantum statistics 
(para-statistics)~\cite{Green, Greenberg, Nelson, Macrae,KD}.

The ``creation and annihilation operators'' for parabosons or parafermions satisfy certain
triple relations, which can be considered as their defining relations.
The main object to construct is then the generalization of the boson or fermion Fock space,
i.e.\ the paraboson or parafermion Fock space.
Such a Fock space (with unique vacuum vector)
is characterized by a certain parameter $p$, known as the order of 
parastatistics.
One way to construct the Fock space of order $p$ was already discussed by Green, and
is referred to as the so-called Green ansatz~\cite{Green}. 
The mathematical equivalent is related to finding a proper basis of an irreducible 
constituent of a $p$-fold tensor product~\cite{Greenberg,KD},
in fact a superalgebra tensor product for parabosons~\cite{Quesne,KD}.
The computational difficulties arising here turn out to be very hard,
and did not lead to a complete solution of the problem (constructing a proper
orthonormal basis of the Fock space, with explicit actions of the paraboson
or parafermion operators on this basis).

In a recent paper~\cite{paraboson} this problem of giving a complete construction
of the Fock space of order~$p$ was solved for the paraboson case.
The solution relies on an important observation by Ganchev and Palev~\cite{Ganchev}, who 
showed that the triple relations for $n$ pairs of paraboson operators are the defining relations 
for the orthosymplectic Lie superalgebra $\osp(1|2n)$~\cite{Kac}.
As a consequence, the paraboson Fock space of order $p$ is a certain
infinite-dimensional unitary irreducible representation (unirrep) of $\osp(1|2n)$.
The construction of this representation was performed in~\cite{paraboson},
using group theoretical techniques and in particular
the branching $\osp(1|2n)\supset \sp(2n)\supset \u(n)$. 
The result is a complete description of a proper basis (an orthogonal 
Gelfand-Zetlin-basis) and the explicit action (matrix elements) 
of the paraboson operators in this basis~\cite{paraboson}.

In the present paper we tackle the same problem for the case of parafermions.
It is already known for a long time that the triple relations for $n$ pairs of parafermion
operators are the defining relations for the orthogonal Lie algebra $\so(2n+1)$~\cite{Kamefuchi,Ryan}.
The parafermion Fock space of order~$p$ then corresponds to a unirrep of $\so(2n+1)$,
namely to the finite-dimensional unirrep $W(p)$ with highest weight $(\frac{p}{2}, \frac{p}{2},\ldots,\frac{p}{2})$.
Contrary to infinite-dimensional unirreps of the Lie superalgebra $\osp(1|2n)$, where
little was known, a lot is known about finite-dimensional unirreps of Lie algebras and
in particular of $\so(2n+1)$. 
For example, it is easy to write down the character of $W(p)$ and compute its dimension.
On the other hand, a proper basis of $W(p)$, in which the action of the parafermion
operators can be given explicitly, is not known. 
The celebrated Gelfand-Zetlin (GZ) basis for finite-dimensional irreducible representations of the
orthogonal Lie groups or Lie algebras, given in~\cite{GZ-so}, is in fact not appropriate here.
In the GZ-basis of~\cite{GZ-so} for $\so(2n+1)$, the action is given for a set of $2n$
generators, which are however different from the parafermion operators (and also different from
the Chevalley generators; in fact, the generators in~\cite{GZ-so} are not root vectors). 
The relations between the parafermion operators and the generators
used in~\cite{GZ-so} or the Chevalley generators are highly nonlinear~\cite{P-fermi}, and hence
do not lead to a practical and explicit expression of the parafermion matrix elements.

Previous attempts to solve this problem followed essentially Green's ansatz.
In terms of the Lie algebra $\so(2n+1)$, this amounts to constructing $W(p)$ as one of the
irreducible components in the $p$-fold tensor product $W(1)^{\otimes p}$ of $W(1)$, where $W(1)$ is the
$\so(2n+1)$ unirrep with highest weight $(\frac{1}{2}, \frac{1}{2},\ldots,\frac{1}{2})$, i.e.\
$W(1)$ is the usual fermion Fock space of dimension $2^n$ (the spinor representation of $\so(2n+1)$).
In a number of papers, further attention was given to the actual decomposition
of $W(1)^{\otimes p}$ into its irreducible components~\cite{Govorkov,BG,Palev1975}. 
In~\cite{BG,Palev1975}, the ``vacuum subspaces'' of the irreducible components of
$W(1)^{\otimes p}$ were considered, and shown to correspond to unirreps of the subalgebra $\u(n)$.
These approaches did not lead to solution for the parafermion matrix elements, however.

In the current paper we present an explicit solution to this problem. 
Our solution is based upon group theoretical techniques: 
we use, in particular, the branching $\so(2n+1)\supset \u(n)$,
an induced representation construction,
the known Gelfand-Zetlin basis description for $\u(n)$~\cite{GZ},
$\u(n)$ Clebsch-Gordan coefficients, the method of reduced matrix elements,
and the computational techniques developed in~\cite{paraboson}.
Our main result is a specific orthogonal basis for $W(p)$
described by means of GZ-patterns, and the explicit action (matrix elements) 
of the parafermion operators in this basis.

The structure of the paper is as follows. In section~\ref{sec:Fock} we give
the defining relations for the parafermion operators, and for the parafermion
Fock space $W(p)$. In section~\ref{sec:Lie} we relate the parafermion operators to a
set of generators for the Lie algebra $\so(2n+1)$, and we give a character formula and
a dimension formula for $W(p)$. Section~\ref{sec:Wp} is the core of the paper: using
an induced module construction, we find a proper basis for $W(p)$ and we compute
the explicit action of the parafermion creation and annihilation operators on the
basis vectors. An example is given in section~\ref{sec:example}, and we conclude
the paper with a summary and some final remarks in section~\ref{sec:summary}.

\setcounter{equation}{0}
\section{The parafermion Fock space $W(p)$} \label{sec:Fock}

For a system consisting of $n$ fermions, with creation and annihilation operators $F_i^\pm$ ($i=1,2,\ldots,n$),
the defining relations are written by means of anticommutators:
\begin{equation}
\{ F_i^-, F_j^+ \} =\delta_{ij},\qquad \{ F_i^-, F_j^- \} =\{ F_i^+, F_j^+ \} =0.
\label{FF}
\end{equation}
The $n$-fermion Fock space is defined as a Hilbert space with vacuum vector $| 0\rangle$, with
\begin{equation}
\langle 0|0\rangle=1, \qquad F_i^- |0\rangle = 0, \qquad (F_i^\pm)^\dagger = F_i^\mp
\qquad (i=1,\ldots,n).
\end{equation}
This Hilbert space is finite-dimensional and irreducible under the action of the algebra
spanned by the elements $1, F_i^+, F_i^-$ ($i=1,\ldots,n$), subject to~\eqref{FF}.
A set of orthogonal basis vectors of this space is given by
\begin{equation}
 |\theta_1,\ldots,\theta_n\rangle = (F_1^+)^{\theta_1}\cdots(F_n^+)^{\theta_n} |0\rangle,
 \qquad \theta_1,\ldots,\theta_n\in\{0,1\} .
 \label{basis-n-fermion}
\end{equation}
This Fock space, denoted by $W(1)$, has dimension $2^n$. 
We shall see that it is a certain unirrep of the Lie algebra
$\so(2n+1)$, with highest weight $(\frac{1}{2},\ldots,\frac{1}{2})$ (the spinor representation).

The purpose of this paper is to study Fock representations of a system of $n$ parafermions. 
For such a system, with creation and annihilation operators $f_i^\pm$ ($i=1,\ldots,n$), the
defining relations are usually given by
\[
[ f_j^-,[f_k^+,f_l^-]] = 2\delta_{jk} f_l^-, \quad
[ f_j^-,[f_k^+,f_l^+]] = 2\delta_{jk}f_l^+ - 2\delta_{jl}f_k^+, \quad
[ f_j^-, [f_k^-, f_l^-]] =0,
\]
and their conjugates. They can be written in the following unified form:
\begin{equation}
[[f_{ j}^{\xi}, f_{ k}^{\eta}], f_{l}^{\epsilon}]=\frac 1 2
(\epsilon -\eta)^2
\delta_{kl} f_{j}^{\xi} -\frac 1 2  (\epsilon -\xi)^2
\delta_{jl}f_{k}^{\eta},  
\label{n-parafermion}
\end{equation}
where $j,k,l\in\{1,2,\ldots,n\}$ and $\eta, \epsilon, \xi \in\{+,-\}$ (to be interpreted as $+1$ and $-1$
in the algebraic expressions $\epsilon -\xi$ and $\epsilon -\eta$).

The parafermion Fock space $W(p)$ is the Hilbert space with unique vacuum vector $|0\rangle$, 
defined by means of ($j,k=1,2,\ldots,n$)
\begin{align}
& \langle 0|0\rangle=1, \qquad f_j^- |0\rangle = 0, \qquad (f_j^\pm)^\dagger = f_j^\mp,\nn\\
& [f_j^-,f_k^+] |0\rangle = p\,\delta_{jk}\,|0\rangle,
\label{pFock}
\end{align}
and by irreducibility under the action of the algebra spanned by
the elements $f_j^+$, $f_j^-$ ($j=1,\ldots,n$), subject
to~\eqref{n-parafermion}. The parameter $p$ is known as the order of the parafermion system.
For $p=1$ the parafermion Fock space $W(p)$ coincides with the fermion Fock space $W(1)$. 
In order to understand the structure of $W(p)$ for general (integer) $p$-values, it will be
necessary to make the connection between the algebraic relations~\eqref{n-parafermion} and
the defining relations of the Lie algebra $\so(2n+1)$.

\setcounter{equation}{0}
\section{The Lie algebra $\so(2n+1)$ and its unirrep $W(p)$} \label{sec:Lie}

The orthogonal Lie algebra $\so(2n+1)$ consists of matrices of the form:
\begin{equation}
\left(\begin{array}{ccc} a&b&c  \\
d&-a^t&e\\
-e^t&-c^t&0
\end{array}\right),
\label{so(2n+1)}
\end{equation}
where $a$ is any $(n\times n)$-matrix, $b$ and $d$ are antisymmetric $(n\times n)$-matrices, and
$c$ and $e$ are $(n\times 1)$-matrices ($t$ stands for transpose). 
Denote by $e_{ij}$ the matrix with zeros everywhere except
a $1$ on position $(i,j)$ (where $i,j\in\{1,\ldots,2n+1\}$). 
Then the Cartan subalgebra $\h$ of $\so(2n+1)$ is 
spanned by the diagonal elements
\begin{equation}
h_j = e_{jj}-e_{n+j,n+j} \qquad (j=1,\ldots,n).
\label{h_j}
\end{equation}
In terms of the dual basis $\epsilon_j$ of $\h^*$ the root vectors and corresponding roots of $\so(2n+1)$ 
are given by:
\begin{align}
e_{jk}-e_{k+n,j+n} & \leftrightarrow \epsilon_j -\epsilon_k, \qquad j\neq k=1,\ldots ,n,\nn\\
e_{j,k+n}-e_{k,j+n} & \leftrightarrow \epsilon_j +\epsilon_k, \qquad j<k=1,\ldots ,n,\nn\\
e_{j+n,k}-e_{k+n,j} & \leftrightarrow -\epsilon_j -\epsilon_k, \qquad j<k=1,\ldots ,n,\label{roots}\\
e_{j,2n+1}-e_{2n+1,j+n} & \leftrightarrow \epsilon_j,  \qquad j=1,\ldots ,n,\nn\\
e_{n+j,2n+1}-e_{2n+1,j} & \leftrightarrow  -\epsilon_j , \qquad j=1,\ldots ,n.\nn
\end{align}
The positive roots are given by $\Delta_+=\{ \epsilon_j \ (j=1,\ldots ,n); \epsilon_j -\epsilon_k,
\epsilon_j +\epsilon_k \ (1\leq j<k\leq n)\}$.
Then it is easy to verify that the following multiples of some of the root vectors 
\begin{equation}
f_{k}^+= \sqrt{2}(e_{j, 2n+1}-e_{2n+1,n+j}), \qquad
f_{k}^-= \sqrt{2}(e_{2n+1, j}-e_{n+j,2n+1}) \qquad (k=1,\ldots , n)
\label{f-as-e}
\end{equation}
satisfy the defining triple relations of $n$ parafermions~\eqref{n-parafermion}. 
Furthermore, note that all other root vectors of $\so(2n+1)$ can be obtained by 
considering all possible commutators $[f_j^{\xi}, f_{ k}^{\eta}]$. 
This was one of the main results of~\cite{Kamefuchi,Ryan}:
\begin{theo} 
As a Lie algebra defined by generators and relations, 
$\so(2n+1)$ is generated by the $2n$ elements $f_k^\pm$ subject to the 
parafermion relations~\eqref{n-parafermion}.
\end{theo}

The parafermion operators $f_j^+$ are part of the positive root vectors, and the
$f_j^-$ are part of the negative root vectors.
The generating vector $|0\rangle$ of the parafermion Fock space $W(p)$ 
satisfies ($j,k=1,\ldots,n$)
\begin{equation}
f_j^- |0\rangle = 0, \qquad
[f_j^-,f_k^+] |0\rangle = p\,\delta_{jk}\, |0\rangle.
\end{equation}
Furthermore, it is easy to verify that
\begin{equation}
[f_j^-,f_j^+]=-2 h_j \qquad(j=1,\ldots,n),
\label{ffh}
\end{equation}
from which one deduces $h_j |0\rangle = -\frac{p}{2}\,|0\rangle$.  
Hence we have the following:
\begin{coro}
The parafermion Fock space $W(p)$ is the unitary irreducible representation of
$\so(2n+1)$ with lowest weight $(-\frac{p}{2}, -\frac{p}{2},\ldots, -\frac{p}{2})$.
\end{coro} 

A lot is known about finite-dimensional representations of the Lie algebra $\so(2n+1)$ 
(or of any simple Lie algebra). In particular, dimension and character formulas are available
for such representations. In these formulas, one uses the Weyl group $\cal W$ and the Weyl tool $\rho$.
For the case of $\so(2n+1)$, ${\cal W}=S_n\times \{-1,+1\}^n$, of order $n!\cdot 2^n$
($S_n$ is the symmetric group). And from~\eqref{roots}
one can verify that the Weyl tool is
\[
\rho = \frac{1}{2} \sum_{\alpha\in\Delta_+} \alpha = 
\sum_{j=1}^n \frac{2n+1-2j}{2} \epsilon_j =
(\frac{2n-1}{2}, \ldots, \frac{5}{2},
\frac{3}{2},\frac{1}{2}).
\]
Weyl's character formula for an irreducible representation $V(\Lambda)$ with highest weight $\Lambda$
reads~\cite{Weyl}
\begin{equation}
\ch V(\Lambda) = \sum_{w\in \cal W} \varepsilon(w) e^{w(\Lambda+\rho)} / 
\sum_{w\in \cal W} \varepsilon(w) e^{w(\rho)},
\label{charWeyl}
\end{equation}
where $\varepsilon(w)=\pm1$ is the signature of $w$, and $e$ stands for the formal exponential.
Usually, one denotes $e^{\epsilon_j}$ by $x_j$; then the character is a (symmetric) polynomial
in the variables $x_1, x_2,\ldots, x_n$.

In the current case, $W(p)$ is an $\so(2n+1)$ representation with lowest weight 
$(-\frac{p}{2}, -\frac{p}{2},\ldots, -\frac{p}{2})$. If $p$ is a positive integer, 
this is the lowest weight of a finite-dimensional
irreducible representation with highest weight $\Lambda=(\frac{p}{2}, \frac{p}{2},\ldots, \frac{p}{2})$.
Note that, with $(\epsilon_1-\epsilon_2, \ldots, \epsilon_{n-1}-\epsilon_n, \epsilon_n)$ as
simple root system, the Dynkin labels of this highest weight are $[0,0,\ldots,0,p]$.
Let us compute the character of this representation: splitting the Weyl group into its symmetric group and 
$\{-1,+1\}^n$, the numerator in~\eqref{charWeyl} becomes
\[
\sum_{w\in \cal W} \varepsilon(w) e^{w(\Lambda+\rho)} =
\sum_{w\in S_n} \varepsilon(w)\ w\left( \prod_{j=1}^n (x_j^{p/2+(2n+1-2j)/2} - x_j^{-p/2-(2n+1-2j)/2})\right);
\]
in terms of a determinant, this can be written as
\begin{align}
\sum_{w\in \cal W} \varepsilon(w) e^{w(\Lambda+\rho)} 
&= \det_{1\leq i,j\leq n} \left( x_i^{p/2+(2n+1-2j)/2} - x_i^{-p/2-(2n+1-2j)/2} \right) \nn\\
& = (x_1\cdots x_n)^{-p/2-(2n-1)/2} \det ( x_i^{p+2n-j}-x_i^{j-1}). \nn
\end{align}
Hence, one obtains
\begin{equation}
\ch W(p) = (x_1\cdots x_n)^{-p/2}\; \frac{\det ( x_i^{p+2n-j}-x_i^{j-1})}{\det ( x_i^{2n-j}-x_i^{j-1})}.
\label{char-det}
\end{equation}
The quotient of determinants appearing here has a nice expression in terms of Schur functions $s_\lambda(x)$, 
see~\cite[p.~84, eq.~($2^\prime$)]{Mac}. It reads:
\begin{equation}
\ch W(p) = (x_1\cdots x_n)^{-p/2} \sum_{\ell(\lambda')\leq p} s_\lambda(x_1,\ldots,x_n).
\label{char-s}
\end{equation}
Herein, the Schur functions $s_\lambda(x)$ are, as usual~\cite{Mac}, labeled by a partition 
$\lambda=(\lambda_1,\lambda_2,\ldots)$, where all $\lambda_i$ are nonnegative integers and
$\lambda_1\geq\lambda_2\geq\cdots$. The number of nonzero parts $\lambda_i$ is the {\em length}
of $\lambda$, denoted by $\ell(\lambda)$. As $s_\lambda(x)=s_\lambda(x_1,\ldots,x_n)$ with only $n$
variables $x_i$, only those $\lambda$ with $\ell(\lambda)\leq n$ appear (since for $\ell(\lambda)> n$
one has $s_\lambda(x_1,\ldots,x_n)=0$). In~\eqref{char-s}, $\lambda'$ stands for the {\em conjugate
partition} of $\lambda$ (a partition whose Young diagram~\cite{Mac} is the transpose of the
Young diagram of $\lambda$). Otherwise said, the sum in~\eqref{char-s} is over all partitions
$\lambda$ whose Young diagram fits inside the $(n\times p)$ rectangle.
In what follows, the expansion~\eqref{char-s} will be very relevant.
For a more general context in which identities such as~\eqref{char-s} appear, see~\cite{Okada}
(the expression~\eqref{char-s} appears there as~\cite[Theorem~2.3(1)]{Okada}).
The origin of~\eqref{char-s} goes back to work of Bracken and Green~\cite{BG1}, 
who do not give~\eqref{char-s} explicitly, but obtain it implicitly by describing
the branching $\so(2n+1)\supset \u(n)$ for $W(p)$.

Apart from the character formula, there is also Weyl's dimension formula~\cite{Weyl} for an irreducible 
representation $V(\Lambda)$ with highest weight $\Lambda$:
\begin{equation}
\dim V(\Lambda) = \frac{\prod_{\alpha\in \Delta_+} \langle \Lambda+\rho | 
\alpha\rangle }{\prod_{\alpha\in \Delta_+} \langle \rho | \alpha\rangle }.
\label{dimWeyl}
\end{equation}
Using~\eqref{roots} and $\langle \epsilon_i|\epsilon_j\rangle=\delta_{ij}$, this yields:
\begin{equation}
\dim W(p) = \prod_{1\leq j\leq k \leq n} \frac{p+1+2n-j-k}{1+2n-j-k}
= \prod_{i=0}^{\lfloor(n-1)/2\rfloor} 
\frac{ \genfrac{(}{)}{0pt}{}{p+2n-2i-1}{2n-4i-1} }{ \genfrac{(}{)}{0pt}{}{2n-2i-1}{2n-4i-1} }.
\label{dim-Wp}
\end{equation}
In the last expression with binomial coefficients, the upper limit for $i$ is the largest integer not exceeding
$(n-1)/2$. 

The character and dimension of $W(p)$ give already a lot of information, but not yet what
we really want, namely an explicit orthogonal basis of $W(p)$ and the action of $f_i^\pm$ on
these basis vectors. 
In order to obtain this, we shall follow the induced module construction used in~\cite{paraboson}.

\setcounter{equation}{0}
\section{The construction of $W(p)$} \label{sec:Wp}

{}From~\eqref{roots} it is already clear that the $2n$ elements $f_j^\pm$ are a set of
generators for $\so(2n+1)$. In fact, all basis elements of $\so(2n+1)$ (all root vectors plus
a basis of the Cartan subalgebra) are given by
\begin{equation}
f_j^\pm\ (1\leq j\leq n), \quad
[f_j^\pm, f_k^\pm]\ (1\leq j < k\leq n),\quad 
[f_j^+, f_k^-] \quad(1\leq j,k\leq n).
\label{so2n+1}
\end{equation}
Some relevant subalgebras of $\so(2n+1)$ are easy to describe by means of these
parafermion generators $f_j^\pm$:
\begin{prop}
A basis for the subalgebra $\so(2n)$ of $\so(2n+1)$ is given by the elements
\begin{equation}
[f_j^\pm, f_k^\pm]\ (1\leq j < k\leq n),\quad 
[f_j^+, f_k^-] \quad(1\leq j,k\leq n).
\label{so2n}
\end{equation}
The $n^2$ elements 
\begin{equation}
[f_j^+, f_k^-] \quad(1\leq j,k\leq n)
\label{un}
\end{equation} 
are a basis for the $\so(2n)$ subalgebra $\u(n)$.
\end{prop}
In fact, with $[f_j^+, f_k^-]=2E_{jk}$, the triple relations~\eqref{n-parafermion} imply the
relations $[E_{ij},E_{kl}]=\delta_{jk}E_{il}-\delta_{li}E_{kj}$. In other words, the 
elements $[f_j^+, f_k^-]$ form, up to a factor 2, the standard $\u(n)$ or $\gl(n)$
basis elements. 

So the parafermion generators $f_j^\pm$ highlight the subalgebra chain $\so(2n+1) \supset \so(2n) \supset \u(n)$.
We use here the notation $\u(n)$, algebraically the same as the general linear Lie algebra $\gl(n)$,
but with the conditions $(f_j^\pm)^\dagger=f_j^\mp$ implying that we are dealing with the ``compact form'' $\u(n)$.

The subalgebra $\u(n)$ can be extended to a parabolic subalgebra~\cite{Warner} ${\cal P}$ of $\so(2n+1)$:
\begin{equation}
{\cal P} = \hbox{span} \{ [f_j^+, f_k^-] \ (1\leq j,k\leq n),\ f_j^-\ (1\leq j\leq n),
[f_j^-, f_k^-] \ (1\leq j<k\leq n)  \}.
\label{P}
\end{equation}
Recall that $[f_j^-,f_k^+] |0\rangle = p\,\delta_{jk}\, |0\rangle$, with $[f_j^-,f_j^+]=-2 h_j$.
This means that the space spanned by $|0\rangle$ is a trivial one-dimensional $\u(n)$ module $\C |0\rangle$
of weight $(-\frac{p}{2}, \ldots, -\frac{p}{2})$.
Since $f_j^- |0\rangle =0$,  the module $\C |0\rangle$ can be extended to a one-dimensional ${\cal P}$ module.
Now we can define the Verma module or the induced $\so(2n+1)$ module $\overline W(p)$:
\begin{equation}
 \overline W(p) = \hbox{Ind}_{\cal P}^{\so(2n+1)} \C|0\rangle.
 \label{defInd}
\end{equation}
This is an $\so(2n+1)$ representation with lowest weight $(-\frac{p}{2}, \ldots, -\frac{p}{2})$.
By the Poincar\'e-Birkhoff-Witt theorem~\cite{Dixmier}, it is easy to give a basis for $\overline W(p)$:
\begin{align}
& (f_1^+)^{k_1}\cdots (f_n^+)^{k_n} ([f_1^+,f_2^+])^{k_{12}}  ([f_1^+,f_3^+])^{k_{13}} \cdots 
([f_{n-1}^+,f_n^+])^{k_{n-1,n}} |0\rangle, \label{Wpbasis}\\
& \qquad k_1,\ldots,k_n,k_{12},k_{13}\ldots,k_{n-1,n} \in \Z_+. \nn
\end{align}
Note that $\overline W(p)$ is infinite-dimensional. It is in general not an irreducible
representation of $\so(2n+1)$. Let $M(p)$ be the maximal nontrivial submodule of $\overline W(p)$. Then the
simple module (irreducible representation), corresponding to the parafermion Fock space, is
\begin{equation}
W(p) = \overline W(p) / M(p).
\label{Wp}
\end{equation}
For $\overline W(p)$, it is easy to compute its character. Using the fact that the weight
of $|0\rangle$ is $(-\frac{p}{2}, \ldots, -\frac{p}{2})$, the basis~\eqref{Wpbasis}, and 
the Cauchy identity~\cite{Little}
\begin{equation}
\frac{1}{\prod_{i=1}^n(1-x_i)\prod_{1\leq j<k\leq n}(1-x_jx_k)} = \sum_{\lambda} s_\lambda (x_1,\ldots,x_n)
= \sum_{\lambda} s_\lambda (x) ,
\label{Schur}
\end{equation}
where the sum is over all partitions $\lambda$ and $s_\lambda(x)$ is the Schur symmetric 
function, one finds
\begin{equation}
\ch \overline W(p) = \frac{(x_1\cdots x_n)^{-p/2}}{\prod_{i=1}^n(1-x_i)\prod_{1\leq j<k\leq n}(1-x_jx_k)}
= (x_1\cdots x_n)^{-p/2} \sum_{\lambda} s_\lambda (x).
\label{char-so2n+1}
\end{equation}
These Schur functions $s_\lambda(x)$ are the characters of finite-dimensional $\u(n)$ representations.
Hence this expansion yields the branching to $\u(n)$ of the $\so(2n+1)$
representation $\overline W(p)$.
This gives an elegant possibility to label the basis vectors of $\overline W(p)$.
For each irreducible representation of $\u(n)$ one can use the corresponding
Gelfand-Zetlin basis~\cite{GZ, Baird}. The union of all these GZ basis is then the basis for $\overline W(p)$.
Thus the new basis of $\overline W(p)$ consists of vectors of the form
(the label $p$ is dropped from the notation of the vectors)
\begin{equation}
 |m)\equiv |m)^n\equiv \left|
\begin{array}{lcllll}
 m_{1n} & \cdots & \cdots & m_{n-1,n} & m_{nn}  \\
 m_{1,n-1} & \cdots & \cdots &  m_{n-1,n-1}  &  \\
\vdots & \qdots & & & \\
m_{11} & & & &
\end{array}
\right) 
= \left| \begin{array}{l} [m]^n \\[2mm] |m)^{n-1} \end{array} \right) \;.
\label{mn}
\end{equation}
The top line of this pattern, also denoted by the $n$-tuple $[m]^n$, is any
partition $\lambda$ (consisting of non increasing nonnegative numbers). 
The remaining $n-1$ lines of the pattern will sometimes be denoted by $|m)^{n-1}$.
All $m_{ij}$ in the above GZ-pattern are nonnegative integers,
satisfying the {\em betweenness conditions}
\begin{equation}
m_{i,j+1}\geq m_{ij}\geq m_{i+1,j+1}\qquad (1\leq i\leq j\leq n-1).
\label{between}
\end{equation}
Since the weight of $|0\rangle$ is $(-\frac{p}{2}, \ldots, -\frac{p}{2})$, the weight
of the above vector is determined by
\begin{equation}
h_{k}|m)=\left(-\frac{p}{2}+\sum_{j=1}^k m_{jk}-\sum_{j=1}^{k-1} m_{j,k-1}\right)|m).
\label{hkm}
\end{equation}

Now we use the same technique as in~\cite{paraboson}. 
The triple relations~\eqref{n-parafermion} yield
\[
[[ f_i^+, f_j^-],f_k^+] = 2\delta_{jk} f_i^+.
\]
With the identification $[ f_i^+, f_j^-]=2E_{ij}$ in the standard $\u(n)$ basis,
this is equivalent to the action $E_{ij}\cdot e_k = \delta_{jk} e_i$. 
Hence the set $(f_1^+,f_2^+,\ldots,f_n^+)$ is a standard $\u(n)$ tensor of rank $(1,0,\ldots,0)$. 
So one can attach a unique GZ-pattern with top line $1 0 \cdots 0$ to every $f_j^+$,
corresponding to the weight $+\epsilon_j$. Explicitly:
\begin{equation}
f_j^+ \sim \begin{array}{l}1 0 \cdots 0 0 0\\[-1mm]
1 0 \cdots 0 0\\[-1mm] \cdots \\[-1mm] 0 \cdots 0\\[-1mm] \cdots\\[-1mm] 0 \end{array},
\label{fGZ}
\end{equation}
where the pattern consists of $j-1$ zero rows at the bottom, and the first $n-j+1$ rows are of the form
$1 0 \cdots 0$.
The tensor product rule in $\u(n)$ reads
\begin{equation}
([m]^n) \otimes (1 0\cdots 0) = ([m]^n_{+1}) \oplus ([m]^n_{+2}) \oplus \cdots \oplus([m]^n_{+n})
\label{untensor}
\end{equation}
where $([m]^n) = (m_{1n},m_{2n},\ldots, m_{nn})$ and a subscript $\pm k$ indicates an increase of 
the $k$th label by $\pm 1$:
\begin{equation}
([m]^n_{\pm k}) = (m_{1n},\ldots,m_{kn}\pm 1,\ldots, m_{nn}).
\label{m+k}
\end{equation}
In the right hand side of~\eqref{untensor}, only those components which are still partitions (i.e.\
consisting of nondecreasing integers) survive.

Now a general matrix element of $f_j^+$ can be written as follows~\cite{paraboson}:
\begin{align}
( m' | f_j^+ | m ) & = 
\left( \begin{array}{ll} [m]^n_{+k} \\[1mm] |m')^{n-1} \end{array} \right| f_j^+
\left| \begin{array}{ll} [m]^n \\[1mm] |m)^{n-1} \end{array} \right) \nn\\
& = \left( \begin{array}{ll} [m]^n \\[2mm] |m)^{n-1} \end{array} ; \right.
 \begin{array}{l}1 0 \cdots 0 0\\[-1mm]
1 0 \cdots 0\\[-1mm]  \cdots\\[-1mm] 0 \end{array}  
\left| \begin{array}{ll} [m]^n_{+k} \\[2mm] |m')^{n-1} \end{array} \right)
\times
([m]^n_{+k}||f^+||[m]^n).
\label{mmatrix}
\end{align}
The first factor in the right hand side is a $\u(n)$ Clebsch-Gordan coefficient~\cite{Klimyk,paraboson},
the second factor is a reduced matrix element.
By the tensor product rule, the first line of $|m')$ has to be of the form~\eqref{m+k},
i.e.\ $[m']^n = [m]^n_{+k}$ for some $k$-value.

The special $\u(n)$ Clebsch-Gordan coefficients (CGCs) 
appearing here are well known.
They can be found, e.g.\ in~\cite{Klimyk}. They are expressed by means of
{\em $\u(n)$-$\u(n-1)$ isoscalar factors} and $\u(n-1)$ CGC's, which on their turn are written by means
of $\u(n-1)$-$\u(n-2)$ isoscalar factors and $\u(n-2)$ CGC's, etc. 
The explicit form of the special $\u(n)$ CGCs appearing here is given in Appendix~A of~\cite{paraboson}.

Just as in~\cite{paraboson}, the main problem is now to find expressions for the reduced
matrix elements, i.e.\ for the functions $G_k([m]^n)$, where 
\begin{equation}
G_k([m]^n) = G_k(m_{1n},m_{2n},\ldots,m_{nn}) = ([m]^n_{+k}||f^+||[m]^n),
\label{Gk}
\end{equation}
for arbitrary $n$-tuples of non increasing nonnegative integers $[m]^n=(m_{1n},m_{2n},\ldots,m_{nn})$.
In that case, one can write the explicit actions:
\begin{align}
f_j^+|m) & = \sum_{k,m'} \left( \begin{array}{ll} [m]^n \\[2mm] |m)^{n-1} \end{array}\right. ;
  \begin{array}{l}1 0 \cdots 0 0\\[-1mm]
1 0 \cdots 0\\[-1mm]  \cdots\\[-1mm] 0 \end{array} 
\left| \begin{array}{ll} [m]^n_{+k} \\[2mm] |m')^{n-1} \end{array} \right)
G_k([m]^n) \left|  \begin{array}{ll} [m]^n_{+k} \\[1mm] |m')^{n-1} \end{array} \right), \label{fj+n}\\
f_j^-|m) & = \sum_{k,m'} \left( \begin{array}{ll} [m]_{-k}^n \\[2mm] |m')^{n-1} \end{array}\right. ;
 \begin{array}{l}1 0 \cdots 0 0\\[-1mm]
1 0 \cdots 0\\[-1mm]  \cdots\\[-1mm] 0 \end{array}  
\left| \begin{array}{ll} [m]^n \\[2mm] |m)^{n-1} \end{array} \right)
G_k([m]_{-k}^n) \left|  \begin{array}{ll} [m]^n_{-k} \\[1mm] |m')^{n-1} \end{array} \right). \label{fj-n}
\end{align}
The key calculation to determine the unknown functions $G_k$ is to start from the following action:
\begin{equation}
[ f_n^-, f_n^+ ] |m) = -2h_n |m) = 
(p-2( \sum_{j=1}^n m_{jn}-\sum_{j=1}^{n-1} m_{j,n-1} )) |m).
\label{fnfn}
\end{equation}
Now one can express the left hand side by means of~\eqref{fj+n}-\eqref{fj-n}, using the explicit form of the 
CGCs and isoscalar factors (which are rather simple in the case $j=n$). The result is a complicated system
of coupled recurrence relations for the functions $G_k$.
It is quite surprising that this system of coupled nonlinear recursion equations has
a rather simple solution. Using the relevant boundary conditions, we have been able to solve this 
system of recurrence relations. This task would have been hardly
impossible without the use of Maple. 
Our main computational result is:
\begin{prop}
\label{prop-main}
The reduced matrix elements $G_k$ appearing in the actions of $f_j^\pm$ on vectors
$|m)$ of $\overline W(p)$ are given by:
\begin{align}
& G_{k}(m_{1n}, m_{2n},\ldots,  m_{nn}) 
=\nn\\
&
\left(-\frac{
({\cal E}_n(m_{kn}+n-k)+1)\prod_{j\neq k=1}^{n} (m_{kn}-m_{jn}-k+j)}
{\prod_{j\neq  \frac{k}{2}=1}^{\lfloor n/2 \rfloor}  (m_{kn}-m_{2j,n}-k+2j)
(m_{kn}-m_{2j,n}-k+2j+1)}
\right)^{1/2}  \label{Gkeven}
\end{align}
for $k$ even; and by
\begin{align}
& G_{k}(m_{1n}, m_{2n},\ldots,  m_{nn}) 
=\nn\\
&
\left(\frac{(p-m_{kn}+k-1)
({\cal O}_n(m_{kn}+n-k)+1)\prod_{j\neq k=1}^{n} (m_{kn}-m_{jn}-k+j)}
{\prod_{j\neq  \frac{k+1}{2}=1}^{\lceil n/2 \rceil}  (m_{kn}-m_{2j-1,n}-k+2j-1)
(m_{kn}-m_{2j-1,n}-k+2j)}
\right)^{1/2} \label{Gkodd}
\end{align}
for $k$ odd.
\end{prop}
Herein ${\cal E}$ and ${\cal O}$ are the even and odd functions defined by
\begin{align}
& {\cal E}_{j}=1 \hbox{ if } j \hbox{ is even and 0 otherwise},\nn\\
& {\cal O}_{j}=1 \hbox{ if } j \hbox{ is odd and 0 otherwise}; \label{EO}
\end{align}
where obviously ${\cal O}_j=1-{\cal E}_j$, but it is still convenient to use both notations.
Also, note that products such as $\prod_{j\neq k=1}^{n}$ means ``the product over all $j$-values
running from 1 to $n$, but excluding $j=k$''. The notation $\lfloor a \rfloor$ (resp.\ $\lceil a \rceil$)
refers to the {\em floor} (resp.\ {\em ceiling}) of
$a$, i.e.\ the largest integer not exceeding~$a$ (resp.\ the smallest integer greater than or equal to $a$).

To present all the details of this computational result is unrealistic. 
Just as in~\cite{paraboson}, the proof consist of verifying that all triple relations~\eqref{n-parafermion} hold
when acting on any vector $|m)$. Each such verification leads to an algebraic identity
in the $n$ variables of the partition, $m_{1n},\ldots, m_{nn}$.
In such computations, there are some intermediate verifications: e.g.\ the action
$[f_j^+,f_k^-] |m)$ should leave the top row of the GZ-pattern $|m)$ invariant (since
$[f_j^+,f_k^-]$ belongs to $\u(n)$). In fact, it must give (up to a factor 2) the known action of
the standard $\u(n)$ matrix elements $E_{jk}$ in the classical GZ-basis.

The explicit expressions for the reduced matrix elements give the
action of the generators in the basis of $\overline W(p)$, for arbitrary $p$.
The structure of the maximal submodule $M(p)$ and hence of the irreducible
factor module $W(p)$ is revealed by examining when these matrix elements vanish.
It follows from~\eqref{Gkeven}-\eqref{Gkodd} that the only crucial factor is
\[
(p-m_{kn}+k-1), \qquad (k\ \rm{odd}).
\]
In particular for $k=1$ this factor is $(p-m_{1n})$, and $m_{1n}$ is the largest
integer in the GZ-pattern. Starting from the vacuum vector, with a GZ-pattern 
consisting of all zeros, one can raise the entries in the GZ-pattern by applying
the operators $f_j^+$. However, when $m_{1n}$ has reached the value~$p$ it can no
longer be increased. As a consequence, all vectors $|m)$ with $m_{1n}>p$ belong to
the submodule $M(p)$. This uncovers the structure of $W(p)$:
\begin{coro}
For $p$ a positive integer, the parafermion Fock space $W(p)$ has as a basis
the vectors $|m)$ with $m_{1n}\leq p$. In other words, the top line of $|m)$ is a
partition $\lambda$ with largest part not exceeding $p$, i.e.\ $\ell(\lambda')\leq p$.
As a consequence, it follows from~\eqref{char-so2n+1} that
\[
\ch W(p) = (x_1\cdots x_n)^{-p/2} \sum_{\lambda,\ \ell(\lambda')\leq p} s_\lambda(x_1,\ldots, x_n).
\]
\end{coro}
This last result coincides with~\eqref{char-s}. 
Note that $p$ must be a positive integer in order to have a positive inner product
$(m|m)$ for all possible patterns.

Of course, we did not make these hard computations just to find the character of
$W(p)$, which was already determined by standard techniques. 
Our main purpose was to find a proper orthonormal basis for $W(p)$, in which the action of
the parafermion operators $f_j^\pm$ can be computed explicitly.
This has now been obtained by means of~\eqref{fj+n}-\eqref{fj-n} and~\eqref{Gkeven}-\eqref{Gkodd}.
Let us summarize this result, inserting also the explicit CGCs of~\cite[Appendix~A]{paraboson}.

\begin{theo}
An orthonormal basis for the parafermion Fock space $W(p)$ is given by the vectors $|m)$,
see~\eqref{mn}, with $m_{1n}\leq p$. Its dimension is given by~\eqref{dim-Wp}. 
The action of the Cartan algebra elements of $\so(2n+1)$ is:
\begin{equation}
h_{k}|m)=\left(-\frac{p}{2}+\sum_{j=1}^k m_{jk}-\sum_{j=1}^{k-1} m_{j,k-1}\right)|m), 
\quad (1\leq k\leq n). \label{h_k} \\
\end{equation}
The action of the parafermion operators $f_j^\pm$ is given below, first for $j=n$ (simple case)
and then for general~$j$. We have:
\begin{align}
f_{n}^+|m)&=\sum_{i=1}^n
\left( \frac{\prod_{k=1}^{n-1}  
(l_{k,n-1}-l_{in}-1 )}
{ \prod_{k\neq i=1}^{n} (l_{kn}-l_{in})}
\right)^{1/2} G_{i}(m_{1n},m_{2n},\ldots,  m_{nn})
|m)_{+in}; \label{fnp} \\[2mm]
f_{n}^-|m)&=\sum_{i=1}^n
\left( \frac{\prod_{k=1}^{n-1}  
(l_{k,n-1}-l_{in} )}
{ \prod_{k\neq i=1}^{n} (l_{kn}-l_{in}+1)}
\right)^{1/2} G_{i}(m_{1n},\ldots, m_{in}-1,\ldots,  m_{nn})
|m)_{-in}; \label{bnm} 
\end{align}
Herein, $l_{ij}=m_{ij}-i$, $G_i$ is determined by~\eqref{Gkeven}-\eqref{Gkodd}, 
and $\pm i,n$ attached as a subscript to $|m)$ indicates a
replacement $m_{i,n}\rightarrow m_{i,n}\pm 1$.
In general,
\begin{align}
f_{j}^+|m)&=\sum_{i_n=1}^n\sum_{i_{n-1}=1}^{n-1}\ldots \sum_{i_j=1}^j
S(i_n,i_{n-1}) S(i_{n-1},i_{n-2})\ldots S(i_{j+1},i_{j})
\left( \frac{\prod_{k= 1}^{j-1}  
(l_{k,j-1}-l_{i_{j},j}-1 ) }
{ \prod_{k\neq i_{j}=1}^{j} (l_{kj}-l_{i_{j},j})}
\right)^{1/2} 
  \nn\\
&\times \prod_{r=1}^{n-j}
\left( \frac{\prod_{k\neq i_{n-r}=1}^{n-r}  
(l_{k,n-r}-l_{i_{n-r+1},n-r+1}-1 )\prod_{k\neq i_{n-r+1}=1}^{n-r+1}(l_{k,n-r+1}-l_{i_{n-r},n-r})
 }{ \prod_{k\neq i_{n-r+1}=1}^{n-r+1} (l_{k,n-r+1}-l_{i_{n-r+1},n-r+1})\prod_{k\neq i_{n-r}=1}^{n-r}(l_{k,n-r}
 -l_{i_{n-r},n-r}-1)}
\right)^{1/2}  \nn \\
&\times G_{i_n}(m_{1n},m_{2n},\ldots,  m_{nn}) \;
|m)_{+i_n,n;+i_{n-1},n-1;\ldots ;+i_j,j} ; \label{Afj+} \\[2mm]
f_{j}^-|m)&=\sum_{i_n=1}^n\sum_{i_{n-1}=1}^{n-1}\ldots \sum_{i_j=1}^j
S(i_n,i_{n-1}) S(i_{n-1},i_{n-2})\ldots S(i_{j+1},i_{j}) 
\left( \frac{\prod_{k= 1}^{j-1}  
(l_{k,j-1}-l_{i_{j},j} ) }
{ \prod_{k\neq i_{j}=1}^{j} (l_{kj}-l_{i_{j},j}+1)}
\right)^{1/2} 
 \nn\\
&\times \prod_{r=1}^{n-j}
\left( \frac{\prod_{k\neq i_{n-r}=1}^{n-r}  
(l_{k,n-r}-l_{i_{n-r+1},n-r+1} )\prod_{k\neq i_{n-r+1}=1}^{n-r+1}(l_{k,n-r+1}-l_{i_{n-r},n-r}+1)
 }{ \prod_{k\neq i_{n-r+1}=1}^{n-r+1} (l_{k,n-r+1}-l_{i_{n-r+1},n-r+1}+1)\prod_{k\neq i_{n-r}=1}^{n-r}(l_{k,n-r}
 -l_{i_{n-r},n-r})}
\right)^{1/2} \nn \\
&\times  G_{i_n}(m_{1n},\ldots,m_{i_n,n}-1,\ldots,  m_{nn}) \;
|m)_{-i_n,n;-i_{n-1},n-1;\ldots ;-i_j,j} . \label{Afj-}
\end{align}
Once again, each symbol $\pm i_k,k$ attached as a subscript to $|m)$ indicates a
replacement $m_{i_k,k}\rightarrow m_{i_k,k}\pm 1$, and
\begin{equation}
S(k,l) = \left\{ \begin{array}{lll}
 {\;\;1} & \hbox{for} & k\leq l  \\ 
 {-1} & \hbox{for} & k>l .
 \end{array}\right.
\end{equation}
\end{theo}

\setcounter{equation}{0}
\section{Example: $W(p)$ for $\so(5)$} \label{sec:example}

The above formulas for general $n$ look rather involved, so it is perhaps useful to give
an example. Let us consider the case $n=2$, i.e.\ the Lie algebra $\so(5)$. For any positive
integer~$p$, the parafermion Fock space $W(p)$ has dimension given by~\eqref{dim-Wp},
\begin{equation}
\dim W(p)= \frac{(p+3)(p+2)(p+1)}{6}.
\label{dim-so5}
\end{equation}
The set of orthonormal basis vectors of $W(p)$ is given by all vectors with integer $m$-patterns of the form
\begin{equation}
\left| \begin{array}{l} m_{12}, m_{22} \\ m_{11} \end{array} \right), \qquad
p\geq m_{12}\geq m_{11} \geq m_{22} \geq 0.
\end{equation}
Clearly, the number of different $m$-patterns satisfying these inequalities coincides with~\eqref{dim-so5}.
The action of the parafermion generators $f_1^+$ and $f_2^+$ is given by:
\begin{align}
f_1^+ \left| \begin{array}{l} m_{12}, m_{22} \\ m_{11} \end{array} \right)  
=& \sqrt{(m_{11}-m_{22}+1)(p-m_{12})}\;  
\left| \begin{array}{l} m_{12}+1, m_{22} \\ m_{11}+1 \end{array} \right) \nn\\
 & -\sqrt{(m_{12}-m_{11})(m_{22}+1)}\;  
\left| \begin{array}{l} m_{12},m_{22}+1 \\ m_{11}+1 \end{array} \right),\label{sol1}\\
f_2^+ \left| \begin{array}{l} m_{12}, m_{22} \\ m_{11} \end{array} \right) 
=& \sqrt{(m_{12}-m_{11}+1)(p-m_{12})}\;  
\left| \begin{array}{l} m_{12}+1, m_{22} \\ m_{11} \end{array} \right) \nn\\
 & +\sqrt{(m_{11}-m_{22})(m_{22}+1)}\;  
\left| \begin{array}{l} m_{12},m_{22}+1 \\ m_{11} \end{array} \right).\label{sol2}
\end{align}
The action of $f_1^-$ and $f_2^-$ follows from the above expressions using $(m'|f_j^-|m) = (m|f_j^+|m')$.
In this case, it is not too difficult to check these actions by hand:
one can do this by comparing the action of the left hand side of the triple relation~\eqref{n-parafermion}
with the action of the right hand side of~\eqref{n-parafermion} (for some values of $j,k,l$ and $\xi,\eta,\epsilon$).

\setcounter{equation}{0}
\section{Summary and conclusion} \label{sec:summary}

In this paper we have given a complete description of the unitary irreducible
representations $W(p)$ of $\so(2n+1)$ with highest weight $(\frac{p}{2},\frac{p}{2},\ldots,\frac{p}{2})$,
which in particular are of interest in the theory of parafermion statistics
because these representations are exactly the Fock spaces of $n$ parafermions,
with $p$ being the order of the parastatistics. 
Although many aspects of $W(p)$ are known from general Lie algebra representation theory,
the problem of giving an appropriate orthogonal basis for $W(p)$ and the explicit action of the
parafermion operators on such a basis was not solved.
In the present paper we solve this problem using some group theoretical methods and computational techniques.
A crucial role in our analysis is played by the $\u(n)$
subalgebra of $\so(2n+1)$, generated by all commutators of the parafermion
creation and annihilation operators. Taking a certain parabolic subalgebra
${\cal P}$ containing $\u(n)$ and a trivial module of ${\cal P}$ generated from
the vacuum, i.e.\ the lowest weight vector of weight $(-\frac{p}{2},-\frac{p}{2},\ldots,-\frac{p}{2})$,
an induced module $\overline W(p)$ of $\so(2n+1)$ is constructed. The Fock module $W(p)$ is the
quotient of this induced module by its maximal submodule $M(p)$. The
character of the induced module is readily obtained and by a classical
result of Cauchy-Littlewood the characters can be rewritten as an
infinite sum over partitions of Schur symmetric functions. This can be
reinterpreted as a decomposition of the $\so(2n+1)$ module into an infinite
sum of finite-dimensional simple $\u(n)$ modules labeled  by partitions. For
each irreducible representation of $\u(n)$ one can use the corresponding
Gelfand-Zetlin basis. The union of all these GZ basis vectors is the basis for
the induced module $\overline W(p)$. The main calculation is then the action
of the parafermion operators on this basis. Just as for the case of
paraboson operators~\cite{paraboson}, the
collection of the $n$ parafermion creation operators is a $\u(n)$ tensor operator of
rank $(1,0,\ldots, 0)$. To calculate its matrix elements, they are written as a product
of certain $\u(n)$ Clebsch-Gordan coefficient and a reduced
matrix element. As the relevant $\u(n)$ CGCs are known, the problem is to find
the reduced matrix elements. Solving a set of recurrence relations for these, leads
to their expressions. These explicit expressions give not only the
action of the generators in the basis of $\overline W(p)$, they also yield
the structure of the maximal submodule $M(p)$ and hence of the irreducible factor module
$W(p)$. This leads to the main result of the paper: an explicit basis of $W(p)$ (consisting
of all possible GZ-patterns with integer entries at most $p$) and the explicit
action of the generators in this basis. As an illustration, the case $n=2$ is given
in more detail.

We have considered here the parafermion Fock spaces of order~$p$ for a finite degree of freedom ($n$ finite).
The real interest lies in such quantum systems (parabosons and parafermions)
with an infinite degree of freedom ($n=+\infty$),
see e.g.~\cite{Greenberg,KD}.
In a forthcoming paper, we hope to report on how the results of~\cite{paraboson}
and of the current paper can be used to construct representations of order $p$
for an infinite set of parabosons and parafermions.
Also an investigation of representations of the ``parastatistics algebra'' in which
both parabosons and parafermions appear~\cite{Loday} should be of interest.

\section*{Acknowledgments}
The authors would like to thank Professor T.D.~Palev and Dr.\ S.\ Lievens for their interest.
N.I.~Stoilova was supported by a project from the Fund for Scientific Research -- Flanders (Belgium)
and by project P6/02 of the Interuniversity Attraction Poles Programme (Belgian State -- 
Belgian Science Policy).


\begin{thebibliography}{99}

\bibitem{Baird}
Baird, G.E., Biedenharn, L.C.: 
On the Representations of the Semisimple Lie Groups. II. 
J.\ Math.\ Phys. {\bf 4}, 1449-1466 (1963)

\bibitem{BG1}
Bracken, A.J., Green, H.S.:
Algebraic Identities for Parafermi Statistics of Given Order.
Nuovo Cim.\ {\bf 9}, 349-365 (1972)

\bibitem{BG}
Bracken, A.J., Green, H.S.:
Parastatistics and quark-model,
J.\ Math.\ Phys. {\bf 14}, 1784-1793 (1973)

\bibitem{Dixmier}
Dixmier, J.:
Enveloping algebras. 
Graduate Studies in Mathematics 11. 
American Mathematical Society, Providence, RI (1996)

\bibitem{Haag}
Dr\" uhl, K., Haag, R., Roberts, J.E.: 
On parastatistics. 
Comm.\ Math.\ Phys. {\bf 18}, 204-226 (1970)

\bibitem{Ganchev}
Ganchev, A.Ch.,   Palev, T.D.:  
A Lie Superalgebraic Interpretation of the Para-Bose Statistics. 
J.\ Math.\ Phys. {\bf 21}, 797-799 (1980)

\bibitem{Green}
Green, H.S.: 
A Generalized Method of Field Quantization.
Phys.\ Rev. {\bf 90}, 270-273 (1953)

\bibitem{GZ}
Gel'fand, I.M.,  Zetlin, M.L.: 
Finite-Dimensional Representations of the Group of Unimodular Matrices.
Dokl.\ Akad.\ Nauk SSSR {\bf 71}, 825-828 (1950)
  
\bibitem{GZ-so}
Gel'fand, I.M.,  Zetlin, M.L.: 
Finite-dimensional representations of groups of orthogonal matrices. 
Dokl.\ Akad.\ Nauk SSSR {\bf 71}, 1017-1020 (1950)
  
\bibitem{Govorkov}
Govorkov, A.B.:
Possibility of a Parafield Representation of Internal Degrees of Freedom like Isospin and Strangeness.
Soviet Phys.\ JETP {\bf 27}, 960-966 (1968)

\bibitem{Greenberg}
Greenberg, O.W.,  Messiah, A.M.L.: 
Selection Rules for Parafields and the Absence of Para particles in Nature.
Phys.\ Rev.\ B {\bf 138} , 1155-1167 (1965)

\bibitem{Nelson} 
Greenberg, O.W.,  Nelson, C.A.: 
Color Models of Hadrons.  
Phys.\ Rep.\ C {\bf  32}, 69-121 (1977)
  
\bibitem{Macrae} 
Greenberg,\ O.W., Macrae,\ K.I.: 
Locally Gauge-Invariant Formulation of Parastatistics.
Nucl.\ Phys.\ B {\bf  219}, 358-366 (1983)

\bibitem{Kac}
Kac, V.G.: 
Lie Superalgebras. 
Adv.\ Math. {\bf 26}, 8-96 (1977)

\bibitem{Kamefuchi}
Kamefuchi, S., Takahashi, Y.:
A generalization of field quantization and statistics.
Nucl.\ Phys. {\bf 36}, 177-206 (1962)

\bibitem{KD}
Kanakoglou, K., Daskaloyannis, C.:
A braided look at Green ansatz for parabosons.
J.\ Math.\ Phys. {\bf 48}, 113516, 19pp (2007)

\bibitem{paraboson}
Lievens, S., Stoilova, N.I., Van der Jeugt, J.:  
The paraboson Fock space and unitary irreducible representations of the Lie superalgebra $\osp(1|2n)$.
Commun.\ Math.\ Phys.\ (in press);	arXiv:0706.4196v2 [hep-th]. 

\bibitem{Little}
Littlewood, D.E.:	
The theory of Group Characters and Matrix Representations of Groups. 
Oxford University Press, Oxford	(1950)

\bibitem{Loday}
Loday, J.-L., Popov, T.:
Parastatistics algebra and super semistandard Young tableaus.
arXiv:0711.3648v1 [math-ph].

\bibitem{Mac}
Macdonald, I.G.: 
Symmetric Functions and Hall Polynomials. Oxford University Press, Oxford, 2nd edition (1995)    

\bibitem{Wu}
Mansouri, F., Wu, Xi Zeng: Parastatistics and Conformal Field Theories in Two Dimensions.
J.\ Math.\ Phys. {\bf 30}, 892-901 (1989)

\bibitem{Okada}
Okada, S.:
Applications of Minor Summation Formulas to Rectangular-Shaped Representations of Classical Groups.
J.\ Algebra {\bf 205}, 337-367 (1998)

\bibitem{Ohnuki}
Ohnuki, Y., Kamefuchi, S.: Quantum Field Theory and Parastatistics. Springer, Berlin (1982)

\bibitem{Palev1975}
Palev, T.D.:
Vacuum-like state analysis of the representations of the para-Fermi operators.
Ann.\ Inst.\ H.\ Poincar\'e Sect.\ A {\bf 23}, 49-60 (1975)

\bibitem{P-fermi} 
Palev, T.D.:
Quantization of $U_q[so(2n+1)]$ with deformed para-Fermi operators.  
Lett.\ Math.\ Phys. {\bf 31}, 151-157 (1994).

\bibitem{Quesne}
Quesne, C.:
Interpretation and extension of Green's ansatz for paraparticles.
Phys.\ Lett.\ A {\bf 260}, 437-440 (1999).

\bibitem{Ryan}
Ryan, C., Sudarshan, E.C.G:
Representations of parafermi rings.
Nucl.\ Phys.\ {\bf 47}, 207-211 (1963)

\bibitem{Klimyk}
Vilenkin, N.Ja, Klimyk, A.U.:  Representation of Lie Groups and Special Functions, Vol. 3:
Classical and Quantum Groups and Special Functions.  
Kluwer Academic Publishers (1992)

\bibitem{Weyl}
Weyl, H.:  The Classical Groups. Princeton, NJ: Princeton University Press (1946)

\bibitem{Warner}
Warner, G.:
Harmonic Analysis on Semi-simple Lie groups I:
Springer, Berlin (1972)

\end{thebibliography}
\end{document}